\documentclass[prd,nofootinbib]{revtex4-1}
\usepackage{latexsym}
\usepackage{amsthm} 
\usepackage{amsmath}
\usepackage{graphicx}
\usepackage{subcaption}
\usepackage{amssymb}
\usepackage{hyperref}
\usepackage{color}
\usepackage{bbold}
\usepackage{simplewick}

\newcommand{\gsim}{\lower.7ex\hbox{$\;\stackrel{\textstyle>}{\sim}\;$}}
\newcommand{\lsim}{\lower.7ex\hbox{$\;\stackrel{\textstyle<}{\sim}\;$}}

\def\mpl{M_{\rm Pl}}
\newcommand{\be}{\begin{equation}}
\newcommand{\ee}{\end{equation}}
\newcommand{\bea}{\begin{eqnarray}}
\newcommand{\eea}{\end{eqnarray}}

\newcommand{\comment}[1]{}
\newcommand{\expect}[1]{\left\langle #1 \right\rangle}

\def\d{\partial}

\def\geff{g_{\rm eff}}

\def\x{\boldsymbol{x}}

\def \k {\boldsymbol{k}}

\def \p {\boldsymbol{p}}
\def \q {\boldsymbol{q}}

\def \md {\mathrm{d}}
\def \bfQ {\textbf{Q}}

\definecolor{summersky}{cmyk}{0.71,0.33,0,0.14}
\definecolor{flamingo}{cmyk}{0,0.51,0.71,0.14}
\def \bfx{\boldsymbol{x}}
\def \vpi {\varphi}
\def \ra {\rangle}
\def \la {\langle}
\def \bfq {{\boldsymbol{q}}}
\def \bfk {{\boldsymbol{k}}}
\def \bp {{\boldsymbol{p}}}
\def \cA {{\cal A}}

\def \geff{g_{\mathrm{eff}}}

\begin{document}

\title{{\Large Cherenkov in the Sky:}
\\
 Measuring the Sound Speed of Primordial Scalar Fluctuations}
\author{Ali Akbar Abolhasani}
\affiliation{Department of Physics, Sharif University of Technology}
\author{Sadra Jazayeri}
\affiliation{DAMTP, Centre for Mathematical Sciences, University of Cambridge, CB3 0WA, United Kingdom}

\begin{abstract}
We consider production of additional relativistic particles coupled to the inflaton. We show that the imprints of these particles on the spectrum of primordial perturbations can be used for the \emph{direct} measurement of the speed of sound of scalar perturbations, regardless of the mechanism of the production of this species. We study a model where these relativistic localized sources are decay products of heavier particles generated via a resonance mechanism. These particles emit in-phase inflaton particles which interfere constructively on the the so-called sound boom, leading to an ``inflationary Cherenkov effect". The resulting shock waves lead to distinctive patterns on the temperature anisotropies of the cosmic microwave background. Moreover, we show that the model predicts unique features on the power spectrum of curvature perturbation and sizeable flattened non-Gaussianity for a suitable range of parameters.
\end{abstract}

\maketitle

\vskip 1 cm


\section{Introduction}
In the simplest scenarios of the cosmic inflation, the universe, at the very onset of its evolution, is almost empty filled with a single scalar field, aka inflaton, whose background energy density is responsible for the nearly exponential expansion of the space\cite{Sato:1980yn,Guth:1980zm,Starobinsky:1980te,Linde:1981mu,Albrecht:1982wi}. On top of this expanding background, quantum fluctuations of this scalar field redshifts to superHubble scales and provide a compelling description for the origin of primordial fluctuations. Generically, these scenarios  give rise to an almost scale invariant, Gaussian and adiabatic spectrum for scalar perturbations consistent with current observations \cite{Akrami:2018odb,Ade:2015ava}. The effective field theory of single field models of inflation has been well developed over the past years \cite{Cheung:2007st, Cheung:2007sv, Senatore:2009gt,Senatore:2010jy,Bartolo:2010bj,Bartolo:2010di,Abolhasani:2015cve}. In this approach, adding a quadratic operator of the form $(g^{00}+1)^2$ leads to a non-standard kinetic term, hence a reduced speed of sound ($c_s<1$). Models with non-standard kinetic term are motivated from high energy physics, in particular low energy effective theories originating from string theory allow for a subluminal speed of sound for the inflaton\cite{ArmendarizPicon:1999rj,Garriga:1999vw, Silverstein:2003hf,Alishahiha:2004eh,Chen:2004gc}. Generally, these models  predict large non-Gaussianity, $f_{NL}\sim 1/c_s^2$, which are very close to equilateral in shape, and also a lower tensor-to-scalar ratio \cite{Cheung:2007st,Akrami:2018odb,Senatore:2009gt}. In this sense, we can only constrain these models by a combined measurement of the power spectrum and the non-Gaussianity parameter $f_{NL}$ \cite{Ade:2015ava,Senatore:2009gt}. However, in the absence of any detection of non-Gaussianity or the primordial tensor pertubrations, recent observations only put mild constraints on the value of the sound speed \cite{Ade:2015ava, Ade:2015tva}. The most stringent bound on $c_s$ is \cite{Akrami:2018odb,Cheung:2007st}
\be   
c_s>0.024\,,
\ee
inferred from the observational bound on primordial non-Gaussianity. The main purpose of this work is to establish a set up for a \emph{direct measurement} of $c_s$ through the inflaton emission by relativistic localized sources during inflation.   

 Our model implicitly relies upon a mechanism that produces super heavy particles as the progenitors of the proposed highly relativistic particles. The phenomenon of particle production during inflation is extensively discussed in the literature; for an incomplete list of references see \cite{Chung:1999ve, Barnaby:2009dd, Barnaby:2010ke, Romano:2008rr,Barnaby:2011qe, Mukohyama:2014gba, Garcia:2019icv, Amin:2015ftc}. In an interesting previous work, it has been shown that particles which are never lighter than $\dot{\phi}^{1/2}$, hence always heavier than Hubble, can be abundantly created during inflation \cite{Flauger:2016idt}. In this scenario, the additional heavy field is coupled to the inflaton via discrete shift symmetric interactions resulting in their \textit{non-adiabatic} production. We assume that these super heavy particles, represented by $\chi$, eventually decay into much lighter species, represented by $A$, within a half-life much shorter than a Hubble time. Moreover, because the decay width of these super massive particles, i.e. $\Gamma\sim M$, is much bigger than Hubble, the secondary particles can be thought as  localized objects with a spatial extension $\Delta x$ much smaller than the Hubble horizon. 
 
 Furthermore, we consider a natural coupling between the relativistic particles and the scalar sector in an EFT framework. The relativistic particles are supersonic, therefore, as they propagate through space they emit inflatons that are coherent in phase, as it happens in the well-known Cherenkov effect. This will lead to shock waves behind each particle's trajectory. The intersection between such shock waves and the last scattering surface can be directly looked for in the CMB anisotropies. In addition, statistical signatures of this secondary inflaton emission would be present in the correlation functions of scalar perturbations as well. The characteristics of both of theses features are sensitive to the value of the speed of sound. 

Let us summarize our main findings,  
\begin{itemize}
\item[$\bullet$]
Based on \cite{Flauger:2016idt}, we imagine a sequence of particle production events, and subsequent decays to relativistic pairs, taking place at the conformal times $\eta=\eta_n$ ($n\in \mathbb{Z}$). As long as successive events are  incoherent, the contributions of each to the scalar power simply add up. The ratio of the scalar emission by relativistic pairs to the  power of the vacuum fluctuations is given by 
\begin{align}
\label{Power-ratio}
\frac{\expect{\zeta(\k,0) \zeta(-\k,0)}'_{\text{Cher}}}{P_{\rm vac}(k)} &= \pi c_s \geff^2 \,\dfrac{ H^2}{4M^2}\,\dfrac{n_{A}}{ H^3 }\, \sum_{n} \left(1-\dfrac{\sin \,2c_s k \eta_n}{2c_s k \eta_n} \right),\qquad (-k \eta_n) <\sqrt{\dfrac{M}{H}}
\end{align}
where $\zeta$ is the curvature perturbations, $M$ is the mass of the progenitors ($M\gg H$), $n_A$ is the physical number density of relativistic particles at the instant of creation, and $g_{\text{eff}}$ is a strength of coupling between the scalar sector and the relativistic particles. We find that even within a regime where the back-reaction of the generated particles on the background is negligible, the amplitude of correction to the scalar power can be substantial.   

\item[$\bullet$]
Every individual super sonic particle will leave a Cherenkov like footprint on the CMB temperature anisotropies. If the number of generated particles per Hubble time per Hubble patch is small enough, the patterns induced by individual particles would be distinguishable. For a particle that moves along the trajectory $\bfx_A=\hat{q}\eta$, the scalar emission $\zeta(\bfx,0)$ at the end of inflation and for $\bfx$'s that are far enough from the creation point is given by
\be
\label{conic}
\zeta(\x,0)=
~\geff~\dfrac{2A_s^{1/2}~c_s^{1/2}}{(2\pi)q_n}\,\dfrac{1}{\sqrt{(\bfx.\hat{q})^2-(1-c_s^2)\bfx^2}}\,, 
\ee
where $A_s$ the amplitude of the primordial scalar fluctuations, and $q_n$ is the comoving momentum of the particle. 
We see that this result blows up towards the so-called Mach cone, i.e. $\bfx . \hat{q}=\sqrt{1-c_s^2} x$.  
\item[$\bullet$]
We show that the bispectrum is expected to pick around the flattened configuration. We show, even for the case that the contribution of Cherenkov emission  to the power spectrum is negligible, $f_{NL}$ can be of order unity or larger. We find hints towards resonances on top of the mentioned flattened bispectrum,  but a rigorous numerical study is crucial to capture the right amplitude and frequency of the oscillations, which we leave for a future work. 
\end{itemize}

The paper is organized as follows. As a warm-up, we begin with a brief review of the Cherenkov radiation in flat space. In section \ref{Che-Rad:Inflation},
we will study the Cherenkov effect in a dS background, assuming that the point-particle approximation holds. In Appendix A, we extensively discuss why in our set up this approximation is valid, on the basis of  a quantum mechanical in-in calculation. We investigate the statistical properties of the model, in \ref{stat1} and \ref{stat2}, by estimating the power spectrum and the bispectrum (in the flattened shape). The distinct features of rare events on the (coordinate space) curvature perturbation is obtained in \ref{Cherenkov:Real-space}. Finally, we conclude in Sec. \ref{Conclusion:Sec}.  
\\\\
Our notations: dot and prime denote the derivative with respect to the conventional and the conformal time, respectively. We use $\epsilon$ to refer to the slow-roll parameter and $A_s$ is the amplitude of scalar perturbations given by
\be
\nonumber
A_s^{1/2}=\dfrac{1}{2\pi}\dfrac{H^2}{(2M_p^2|\dot{H}|c_s)^{1/2}}\,.
\ee
 Moreover, $\int_{\q}$ is the shorthand for the integration over momentum $\int \frac{d^3 q}{(2\pi)^3}$, and $\langle....\rangle '$ is an $N$-point function in the Fourier space with the factor $(2\pi)^3\delta^3(\bfk_1+..+\bfk_N)$ dropped.  We use the mostly positive signature for the metric and the natural units.
\section{Cherenkov Radiation in Flat Space}
Before considering the Cherenkov radiation of particles during inflation, let us investigate a toy example in flat space (with  coordinates $(\eta,\bfx)$). Suppose that a massless scalar particle prepared in a wave-packet state passes through a medium at the speed of light. Assume that this particle, represented by $A(\eta,\bfx)$, is coupled to another scalar field through the following cubic vertex
\be
{\cal L}_{\mathrm{int.}}= -\dfrac{1}{2}\,g \varphi A^2\,,
\ee
and that $\vpi$ has a non-relativistic dispersion relation, characterized by a speed of sound $c_s<1$.
 
For concreteness, we take the following initial quantum state for the system:
\be
\label{packet}
|\Psi(\eta_n)\ra=|0\ra_{\phi} \otimes \int_{\q}\,{\cal A}(\bfq)\,a_A(\bfq)^\dagger\,|0\ra_{A}\,,
\ee
where $a_A^\dagger$ is the creation operator associated with A  particles, $|0\ra$ represents the vacuum state, and ${\cal A}(q)$ is a sharp Gaussian distribution around a central momentum $\bp$, i.e. 
\be
\cA (\bfq)=\Big(\sqrt{\dfrac{2\pi}{\Delta p^2}}\Big)^{3/2}\, \exp\Big(-\dfrac{(\bfq-\bp)^2}{4\Delta p^2}\Big)\,\exp(iq\eta_n)\,,\quad\quad \dfrac{\Delta p}{p}\ll 1\,.
\ee
The quantities of our primary interest are the equal time correlation functions of $\vpi$ at a later time, say $\eta=0$.  Using the \textit{in-in} formalism, we start off by computing the average of $\vpi$ in Fourier space, i.e. 
\be
\label{inin1}
\la \Psi|\,\vpi_{\k} (0)\,|\Psi\ra\,=-\,\dfrac{g}{2}\int_{\eta_n}^{0}d\eta\,\, 2\text{Im}\,\Big[f_\vpi(k,\eta)f_\vpi(k,0)\Big]\,\la \Psi|\,A^2(\bfk,\eta)\,|\Psi\ra+{\cal O}(g^3)\,.
\ee
Above, $f_\vpi(k,\eta)$ is  the mode function of $\vpi$, i.e.
\be
f_\vpi(k,\eta)=\dfrac{1}{\sqrt{2c_s\,k}}\exp(-i\,c_sk\eta)\,,
\ee
and $2\text{Im}\Big[f_\vpi(k,\eta)f_\vpi(k,0)\Big]$ is the corresponding retarded Green function. The $\la \Psi|A^2(\bfk,\eta)|\Psi\ra$ acts as a source and is given by
\be
\label{Jsou}
J(\bfk,\eta)=\la \Psi|A^2(\bfk,\eta)|\Psi\ra=2\int_\bfq\,\cA(\bfq)^*\,\cA(\bfq-\bfk)\,f_A(|\bfq+\bfk|,\eta)\,f^*_A(q,\eta)\,,
\ee
where $f_A(k,\eta)=\dfrac{1}{\sqrt{2k}}\exp(-i\,k\eta)$ is the A particle mode function. Evidently, for $k\gg \Delta p$ the source is exponentially suppressed. Moreover, an enhancement in $\vpi_\bfk$ takes place if $k$ is \textit{soft}, in the following sense
\be
\label{softness}
k^2\ll \dfrac{p}{\eta_n}\,. 
\ee
In this regime, the two wave functions in \eqref{Jsou} resonate and enhance the integral (see Appendix \ref{ppinin}). Then the following approximation can be used to simplify the expression for $\vpi_\bfk$, 
\be
\label{softap}
\exp\Big(i (q-|\bfq+\bfk|)\eta\Big)\sim \exp(-i\bfk.\hat{q}\eta)\,.
\ee

By assuming \eqref{softness} and \eqref{softap}, we arrive at
\be
\label{1pf}
\vpi_{\text{cl}}\equiv \la \Psi|\,\vpi_\bfk(0)\,|\Psi\ra\,=-\,g\int_{\eta_n}^{0}d\eta\,\text{Im}\,\Big[f(k,\eta)f(k,0)\Big]\,\dfrac{1}{p}\exp\Big(-i\bfk.\hat{p}(\eta-\eta_n)\Big)\,.
\ee
This is precisely the radiation profile of a point-particle that is moving on the world-line $x^\mu=(\eta,c_s(\eta-\eta_n)\hat{p})$, and is coupled to the $\vpi$ field via
\be
{\cal L}_{\text{int}}=g\int d^4x\,\vpi(x)J(x)\,,\qquad\,J(x)=-\int \dfrac{d\eta'}{p^0(\eta')}\delta^4\Big(x^\mu-x^\mu (\eta')\Big)\,,
\ee
where $p^\mu$ is the four-momentum of the particle. In real space, and far away enough from the initial position of the particle, $\vpi$ becomes
\be
\label{cerenkov}
\la\Psi|\vpi(\bfx+\hat{p}\eta_n,\eta=0)|\Psi\ra=-\dfrac{g\,c_s}{2\pi\,p}\dfrac{1}{\sqrt{(\bfx.\hat{p})^2-(1-c_s^2)r^2}}\theta (-(1-c_s^2)^{1/2}-\hat{\bfx}.\hat{p})\,.
\ee
where $r\equiv\sqrt{\bfx^2}$, and the particles lies at the origin at $\eta=\eta_n$. 
The above  has the celebrated conic structure in the Cherenkov phenomenon: the radiated $\vpi$ is only non-zero inside the cone,  and diverges on the Mach cone. 

Quite remarkably, the classical behavior observed in \eqref{1pf}, reoccurs for an arbitrary correlation function of $\vpi$ \cite{Mirbabayi:2014jqa}(see Appendix \ref{ppinin} for a proof). Meaning, as long as the external momenta $\bfk_i$ are soft enough, e.g. $k^2_i\ll \frac{p}{\eta_n}$, a generic n-point function of $\vpi$ can be written as
\be
\label{Npt}
\la \Psi|\,\vpi_{\bfk_1}(0)\,...\,\vpi_{\bfk_N}(0)\,|\Psi\ra=\la 0|\vpi_{\text{eff}}(\bfk_1,0)\,...\,\vpi_{\text{eff}}(\bfk_N,0)\,|0\ra\,,
\ee
in which the effective operator is defined through
\be
\label{classical}
\vpi_{\text{eff}}(\bfk,\eta)=\hat{\vpi}_{\text{vac}}(\bfk,\eta)+\mathbb{1}\,\vpi_{\text{cl}}(\bfk,\eta)\,.
\ee
Above, $\vpi_{\text{cl}}$ is the c-number defined in \eqref{1pf}, $\mathbb{1}$ is the unity operator, and $\hat{\vpi}_{\text{vac}}$ is the Heisenberg operator corresponding to the scalar field. In short, the above equality means that the soft radiation induced by the motion of particle is almost classical. Notice that, as it is the case for the one-point function, correlators exponentially dilute for very hard modes, defined by $|\bfk_1+...+\bfk_N|>\Delta p$, and are power law suppressed for hard modes, i.e. for $k_i\gsim \sqrt{p/\eta_n}$.  

Scenarios of particle production in an expanding universe incorporate \textit{squeezed states} which cannot be written as a direct product of one-particle wave packets. As an illustration,  consider the following state 
\be
\label{sqst}
|\Psi\ra= |0\ra_\vpi\,\otimes\,{\cal N}\,\exp\Big(\int \dfrac{d^3q}{(2\pi)^3}\dfrac{\beta_q}{2\alpha_q}\,a_A(\bfq)^\dagger\,a_A(-\bfq)^\dagger\Big)\,|0\ra\,,
\ee
in which ${\cal N}$ is a normalization factor. $|\Psi\ra$ can be generated by performing a Bogoliubov transformation on the vacuum state of the A sector with coefficients $\alpha_k$ and $\beta_k$ standing in front of the positive and negative frequencies, respectively.
This kind of squeezed states are ubiquitous in cosmology and reheating, in particular (see, for example, \cite{Kofman:1997yn})--- they typically emerge due to non-adiabatic change in the time dependent mass of the particle.  Notice that \eqref{sqst} respects isotropy as $\beta$ and $\alpha$ only depend on the size of the particle momentum. 

We are interested in the limit where the particle description pertains, i.e. we assume $|\beta_q|^2\ll 1$---the occupation number in the phase space is small, and $\beta_q$ sharply peaks around a particular momentum $p$. Similar to the previous computation, the correlation functions of $\vpi$ can be computed with \eqref{sqst} taken as the initial condition. As one might  anticipate, there is a regime where the same classical behavior as before persists: the correlation functions of soft momenta are as if they originated from a  homogeneous distribution of classical point-particles, with the initial number density $n_A=\int_\bfq\,|\beta_q|^2$. In this limit the connected part of the correlation function becomes 
\bea
\label{cersq}
\la \Psi|\,\vpi_{\bfk_1}(0)\,...\vpi_{\bfk_N}(0)|\Psi\ra_{\text{conn.}}&&\sim (2\pi)^3\,\delta^3(\bfk_1+...+\bfk_N)\,n_A\,\\ \nonumber
&&\qquad\quad\int\,\dfrac{d^2\hat{p}}{4\pi}\prod_{i=1}^N\, (-g)\,\int_{\eta_n}^0\,d\eta\,\text{Im}\Big[f(k,\eta)f(k,0)\Big]\,\dfrac{1}{p}\,\exp(-i\bfk_i.\hat{p}\eta)\,.
\eea
A few remarks on this result are in order,
\begin{itemize}
\item Similar to our previous example, the approximation in \eqref{cersq} holds only for $k_i^2\lesssim \frac{p}{\eta_n}$ ---In larger momenta, the amplitude receives suppression of order $1/(p\eta_n)^2$. 
\item The squeezed state taken above respects the translational invariance, hence the conservation of the momenta. Thus, and opposed to \eqref{inin1}, the exponential suppression in the radiation of hard momenta, defined by $k>\Delta p$, disappears.
\item In \eqref{cersq} seemingly leading order ${\cal O}(\beta)$ terms are ignored. These are interference contributions, and are quantum in nature\cite{Flauger:2016idt}. However, they are suppressed by a factor of $\frac{1}{p\eta_n}$, hence we can consistently neglect them. 
\item  The correlation functions in \eqref{cersq} can be interpreted in the same manner as in \eqref{classical}, except that the source of $\vpi_{\text{cl}}$ is now a homogeneous (in space) and isotropic (in momentum) \textit{random distribution} of point-particles. The explicit form of the source is given by 
\be
\label{sourceJ}
J(\bfx,\eta)=\dfrac{g}{p}\sum_{i,j}Y(\bfx_i,\hat{p}_j)\,\delta^3(\bfx-\hat{p}_j(\eta-\eta_n)-\bfx_i)\,.
\ee
In this expression, $i$ labels hypothetical small cells with the small volume $\Delta p^{-3}$ covering the entire space at $\eta=\eta_n$, $\bfx_i$ stands for the coordinate of the $i$th cell, and $j$ refers to the quantized direction of the particles mometum. $Y(\bfx_i,\hat{p}_j)$ represents a random variable which is either 1 or 0, depending on respectively the presence or absence of a particle inside the attributed cell and having the corresponding momentum $p\hat{p}_j$
\footnote{Assuming $|\beta|^2\ll 1$ ensures that in each cell at most one particle can exist.}. (See appendix A of \cite{Mirbabayi:2014jqa}.)
Once coupled to the $\vpi$ sector the source \eqref{sourceJ} induces the following classical solution
\be
\vpi_{\text{cl}}(\bfk,\eta=0)=\sum_{i,j}(-g)\int_{\eta_n}^{0}d\eta\,\text{Im}\,\Big[f(k,\eta)f(k,0)\Big]\,\dfrac{1}{p}\exp\Big(-i\bfk.\hat{p_j}(\eta-\eta_n)\Big)Y(\bfx_i,\hat{p}_j)\,\exp(-i\bfk.\bfx_i)\,.
\ee
Notice that in the limit $k_i\ll  \Delta p$, the translational invariance is recovered, hence the conservation of spatial momentum. 
In addition, in order to correctly reproduce \eqref{cersq} the statistics of $Y$s must be taken as
\be
\label{Ydis}
\langle Y(\bfx_{i_1},\hat{q}_{j_1})\,...\,Y(\bfx_{i_m},\hat{q}_{j_m})\rangle=\Delta p^{-3}\, n_A\,\delta_{i_1\,i_2}\delta_{j_1\,j_2}\,...\,\delta_{i_{m-1}i_m}\delta_{j_{m-1}\,j_m}\,.
\ee
In other words, for each cell $Y_{i,j}=1$ with probability $p=n_A\,\Delta p^{-3}\ll 1$, and is zero with probability $(1-p)$, $Y$'s of different cells are completely uncorrelated, the probability in particle's momentum is isotropic with vanishing correlation between any two nonidentical directions.

\end{itemize}
\section{Cherenkov Radiation During Inflation}
\label{Che-Rad:Inflation}
Although we are interested in the production of relativistic particles in a cosmological setting, at the level of model-building we find it extremely challenging, if not impossible, to generate a squeezed state of (nearly) massless particles directly from the cosmological time-dependent background. On the other hand, scenarios incorporating abrupt production of non-relativistic particles have been frequently studied in the literature. Therefore, it is quite reasonable to consider two independent interactions: one with the time dependent background that eventually results in the production of relativistic sources, and the other, between the relativistic particles and the inflaton, which leads to the Cherenkov effect. To this end, we will be considering scenarios in which, in one or a  series of events, a bunch of massive $\chi$ particles becomes non-adiabatically generated via the resonance mechanism explained in \cite{Flauger:2016idt}, and each particle, with a life time much smaller than one Hubble time, decays into two light particles of the type A . 

From an EFT standpoint, the leading order interaction between A  particles and the Goldstone boson of the time translation $\pi$ should respect the approximate shift symmetry of $\pi$ during the slow-roll phase \cite{Cheung:2007st}. Then, up to the lowest order in derivatives and after taking the decoupling limit, the effective action becomes
\be
\label{eft}
{\cal S}=\int \sqrt{-g}\,d^4x\,\left[\dfrac{-M_p^2\dot{H}}{c_s^2}\,\left(\dot{\pi}^2-\dfrac{c_s^2}{a^2} \partial_i\pi\,\partial_i \pi\right)\,-\dfrac{1}{2}m_A^2\,A^2+\dfrac{1}{2}\,g^2\,\Big( g^{\mu\nu}\partial_\mu(t+\pi)\partial_\nu(t+\pi)-1\Big)\,A^2\right]\,.
\ee 
As one might anticipate, introducing the cosmic expansion does not significantly change the point-particle picture discussed in the previous section. We postpone the justification of the point-particle approximation to Appendix \ref{ppinin}, which is based on a first principle in-in calculation and here we briefly review its main points. First, to avoid any IR secular growth in the two point function of $\pi$ due to particle production loops we should take the $A$ particles to be heavier than $\sqrt{2}H$, this means
\be
\nonumber
m_A\gsim \sqrt{2}H\,.
\ee
Second, a single event occurred at $\eta=\eta_n$ only influences the correlation functions of \textit{soft} $\pi$'s, whose momenta lies withing the following range, 
\be
\label{soft}
k\lesssim k_{\text{UV}}=\dfrac{1}{|\eta_n|}\left(\dfrac{M}{H}\right)^{1/2}\,,
\ee
where $M$ is the mass of the progenitor $\chi$ particle. On the other hand, the impact of the non-Bunch Davies vacuum on the $\pi$ correlation functions with larger momenta would be generically suppressed with one or more powers of $\dfrac{1}{|k\eta_n|}$. 

Within the point-particle approximation, the A  particles interact with $\pi$ through an effective mass that depends on $\pi$ and can be read directly from \eqref{eft}, i.e.
\be
m^{\text{eff}}_A(\pi)=\Big(m_A^2+g^2(2\dot{\pi}-(\d_\mu\,\pi)^2)\Big)^{1/2}\,.
\ee
To leading order in derivatives, the point-particle effective action reads
\be
{\cal S}_{\text{pp}}=-\int d\tau\,m\Big(\pi(x^\mu(\tau))\Big)=-\int\,d\tau\,(m_A+\dfrac{g^2}{m_A}\dot{\pi}+...)\,,\qquad d\tau\equiv \sqrt{-g_{\mu\nu}\,dx^\mu\,dx^\nu}\,,
\ee
 where $\tau$ denotes the proper time along the trajectory of the A particles.

Focusing on the interacting part, we have
\be
\label{ppeff}
{\cal S}_{int} = -\dfrac{g^2}{m_A}\int\,d\tau\,\dot{\pi}(t,\bfx(t))=-g^2\,\int_{\eta_n}\,d\eta\,d^3\bfx\,\pi'(\eta,\bfx)\,\Big(\dfrac{1}{E(\eta)}\,\delta^3(\bfx-\bfx(\eta))\Big)\,,
\ee
where $t$ is replaced with the conformal time $\eta$, prime denotes derivative with respect to $
\eta$, and $E(\eta)$ is the time dependent energy of the particle, i.e.
\be
E(\eta)=m_A\,\gamma(\eta)=m_A\,\left(1+\dfrac{\eta^2}{\eta_n^2}\left(\dfrac{E_n^2}{m_A^2}-1\right)\right)^{1/2}\,,\quad \text{where}\, \quad E_n=E(\eta_n)\,.
\ee
Since A particles are generated through the decay of $\chi$'s, their initial energy is $E_n=-H q_n\eta_n=M/2$, where $q_n$ is their initial comoving momentum. The indice $n$ is adopted to represent the possibility of having numerous events, each having a different $q_n$ proportional to $1/\eta_n$. 

Hereafter, it is easier to work with the canonically normalized $\pi$, defined by
\be
\pi_c=\sqrt{\dfrac{2M_p^2|\dot{H}|}{c_s^2}}~\pi.
\ee
 The classical radiation emitted by a point-particle that moves along a trajectory given by $\bfx_A(\eta)$ can be straightforwardly calculated as below
\be
\pi_c(\bfk,\eta=0 )= \int_{\eta_n}^0 \md \eta\, G_k(\eta) J(\k,\eta)\,. 
\ee
Inside the integrand, $J(\bfk,\eta)$ is the point particle source that appears on the rhs of the Fourier space equation of motion for $\pi_c$, and is given by
\bea
\label{source-term:Jk}
J(\k,\eta)&=& \dfrac{g^2 c_s}{\sqrt{2M_p^2|\dot{H}|}}\,\dfrac{d}{d\eta}\,\Big(\dfrac{1}{E(\eta)}\exp(-i\bfk.\bfx_A(\eta))\,\theta(\eta-\eta_n)\Big)\,,
\eea
moreover, $G_k(\eta)$ is $\pi_c$'s bulk-to-boundary retarded Green function  , i.e.   
\be
G_k(\eta) = 2f_{\pi}(k,0) \,\mathrm{Im} f_{\pi}(k,\eta)\,,
\ee
where $f_{\pi}(k,\eta)$ is the mode function of $\pi_c$ with the Bunch-Davies vacuum normalization
\be 
f_{\pi}(k,\eta) = \dfrac{H}{\sqrt{2 c_s^3 k^3}}(1+ic_s k\eta) e^{-i c_s k\eta}.
\ee
By putting everything together, the scalar emission found to be
\bea\label{zeta-k:step1}
\pi_c(\bfk,\eta=0)=-g_{\mathrm{eff}}\,f_\pi(k,0)\,\int_{\eta_n}^0\, \text{Im}\,f'_\pi(k,\eta)\,\dfrac{1}{E(\eta)}\exp\Big(-i\bfk.\bfx_A(\eta)\Big)\,,
\eea
where the time derivative in \eqref{source-term:Jk} has been integrated by part
\footnote{In dealing with \eqref{source-term:Jk}, avoiding the ambiguity in the value of $\Theta(0)$ demands the time derivative of $\Theta(\eta-\eta_n)$ to be taken prior to the mentioned integration by part, and therefore no boundary term at $\eta=\eta_n$ will appear. The final answer is in agreement with the explicit in-in calculation provided in Appendix \ref{ppinin}.}
, moreover, $\geff$ is the dimensionless coupling between the canonically normalized scalar field $\pi_c$ and the relativistic particle, i.e.
\be
\geff \equiv \dfrac{2g^2c_s}{\Big(2M_p^2|\dot{H}|\Big)^{\frac{1}{2}}}\,.
\ee

To facilitate our analytical study, we assume that the particle is relativistic all the way along its trajectory, so we have 
\be
E(\eta)\sim \dfrac{M}{2}\dfrac{\eta}{\eta_n}=-H\,q_n\,\eta\,,\qquad\text{and}\qquad \bfx_A(\eta)=\bfx_n+\hat{q}(\eta-\eta_n)\,.
\ee

Putting everything together, 
\bea
\label{zeta-k:by-parts}
\dfrac{\pi_c(\bfk,0)}{f_\pi(k,0)}=\geff\dfrac{\sqrt{c_s/2\,}}{q_n}\,e^{-i\bfk.(\bfx_n-\eta_n)}\, g(\bfk,\hat{q},\eta_n)\,,
\eea
where 
\be
\label{gform}
g(\bfk,\hat{q},\eta_n)=k^{1/2}\,\int_{\eta_n}^0\,d\eta\,\sin(c_s\,k\eta)\exp(-i\bfk.\hat{q}\eta)\,.
\ee
In our derivation, $m_A$ has been neglected. In order to estimate the error due to this approximation, notice that retrieving $m_A$, among other things, would cut off the time integral in \eqref{gform} at 
\be
\label{devrel}
 |\eta_{\text{IR}}|\sim \dfrac{m_A}{q_n\,H}\sim |\eta_n| \,\dfrac{m_A}{M} \,,
\ee
which corresponds to the time at which the $A$ particles come to rest. For this IR cut-off to induce negligible effect on the final result \eqref{zeta-k:by-parts}, $m_A$ should be smaller than an upper bound. Replacing $\eta=0$ with $\eta_{\text{nr}}$ generates deviations of order $|k\eta_{\text{IR}}|$ in \eqref{gform}, however $k$ is bounded from above itself, i.e.  
\be
|k\eta_{\text{IR}}|\ll |k_{\text{UV}}\eta_{\text{IR}}|\sim \dfrac{m_A}{\sqrt{M\,H}}\,,
\ee
thus, the effect of ignoring $m_A$ would be negligible as long as
\be
\dfrac{m_A}{\sqrt{M\,H}}\ll 1\,.
\ee
\subsection{Frequent Events: Statistical Properties}
\label{stat1}
In \eqref{zeta-k:by-parts}, having found the radiation of a single particle, calculating that of a stochastic distribution of particles is straightforward. Consider the same initial distribution as in \eqref{Ydis}, and by simply summing over all particles' contributions in \eqref{zeta-k:by-parts}, albeit weighted by the $Y$ random-variable, we arrive at
\bea
\label{pp:2pf}
&&\dfrac{\langle \pi_c(\bfk,0)\pi_c(-\bfk,0)\rangle_{\text{Cher}}}{P_{\text{vac}}}=
\\ \nonumber
 && \qquad\qquad \qquad n_A\, a_n^3\,\int \dfrac{d^2\hat{q}}{4\pi} \left(\geff\,\int_{\eta_n}^0\, \text{Im}\,f'_\pi(k,\eta)\,\dfrac{1}{q_n\,\eta\,H}e^{-i\bfk.\hat{q}\eta}\right)\left(\geff\,\int_{\eta_n}^0\, \text{Im}\,f'_\pi(k,\eta)\,\dfrac{1}{q_n\,\eta\,H}e^{+i\bfk.\hat{q}\eta}\right)
\eea
where $\text{Cher}$ represents the Cherenkov radiation contribution to the two point function, $n_A$ is the physical initial number density of the A particles, $a_n=-\frac{1}{\eta_n\,H}$ is the scale factor at $\eta_n$, and $P_{\text{vac}}=|f_\pi(k,0)|^2$ stands for the vacuum fluctuations of $\pi_c$. In light of the explicit in-in quantum mechanical calculation performed in Appendix \ref{ppinin}, the point-particle approximation, and hence \eqref{pp:2pf}, applies to soft momenta defined by \eqref{soft}. 
For larger momenta, the rhs of \eqref{pp:2pf} is suppressed with $\dfrac{1}{|k\eta_n|^2}$ (see Appendix \ref{ppinin} for discussions). 

\begin{figure}
\begin{center}
\includegraphics[scale=0.75]{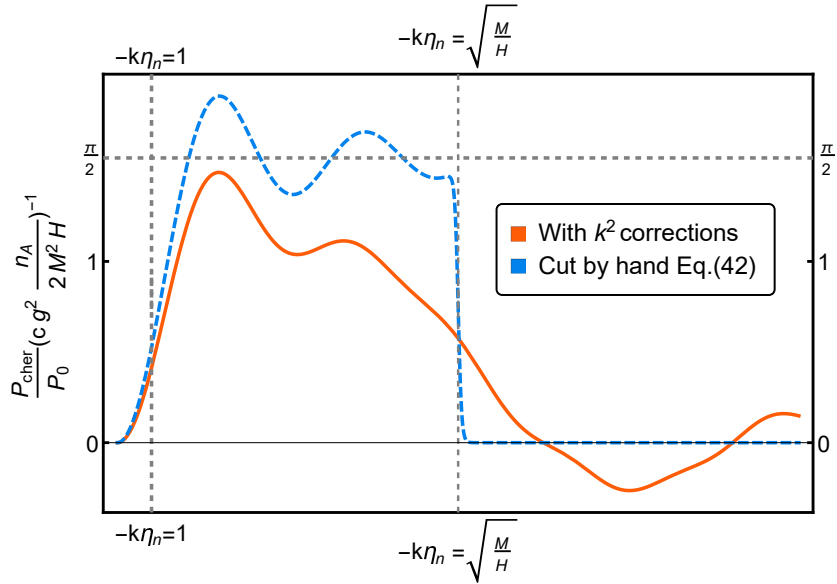}
\caption{\label{Cher-Spec:fig} Power spectrum of scalar perturbations for a single event vs $k |\eta_n|$. (a) Blue Dashed line is the Eq.\eqref{Power-ratio:1} where the power is cut by hand at momentum $-k \eta_n =\sqrt{M/H}$. (b) Solid red line depicts the result of numeric integration of Eq.\eqref{2pfds} considering the next to leading order correction in $k$ which shows a $1/(-k\eta_n)^2$ suppression.}
\end{center}
\end{figure}
So far we have computed the contribution of a single event to the two point function, however, it is more realistic to assume that the $A$ s are periodically created, e.g. due to the periodic creation and rapid decay of $\chi$'s. As long as different particle production events are incoherent and $|\beta|\ll 1$, individual contributions to the two point function would linearly sum up \cite{Mirbabayi:2014jqa}. Therefore, we can write 
\begin{align}
\label{Power-1st}
\frac{\expect{\pi_c(\k,0) \pi_c(-\k,0)}'_{\text{Cher}}}{P_{\text{vac}}(k)} &= c_s \geff^2\,\dfrac{n_A}{2M^2\,H } \, \sum_{n} \dfrac{1}{-\eta_n} \int \dfrac{\md \hat{\q}}{4 \pi} |g(\k,\hat{\q},\eta_n)|^2\,.
\end{align}
The integral over $\hat{\q}$ is straightforward 
\begin{align}
\nonumber
\int \dfrac{\md \hat{\q}}{4 \pi} |g(\k,\hat{\q},\eta_n)|^2 &=  k \int^0_{\eta_n} \int^0_{\eta_n} \md \eta_1\,\md \eta_2 \sin c_s k \eta_1\,\sin c_s k \eta_2\, \dfrac{\sin k(\eta_1-\eta_2)}{k(\eta_1-\eta_2)}.
\end{align}
Though the result of the above integral cannot be written in terms of elementary functions, a reasonable estimate can be made by noting that for $|k \eta_n|\gg 1$, the function $\sin k(\eta_1-\eta_2)/(\eta_1-\eta_2)$ is almost the same as $\pi \delta(\eta_1-\eta_2)$. Consequently, we find
\begin{equation}
\boxed{
\label{Power-ratio:1}
\frac{\expect{\pi_c(\k,0) \pi_c(-\k,0)}'_A}{P_{\rm vac}(k)} = \pi c_s \geff^2 \,\dfrac{ H^2}{4M^2}\,\dfrac{n_{A}}{ H^3 }\, \sum_{n} \left(1-\dfrac{\sin \,2c_s k \eta_n}{2c_s k \eta_n} \right),\qquad (-k \eta_n) <\sqrt{\dfrac{M}{H}}
}
\end{equation}
On the opposite limit, i.e. when $|k\eta_n|\ll 1$, Eq. \eqref{Power-1st} simplifies into
\be
\frac{\expect{\pi_c(\k,0) \pi_c(-\k,0)}'_{\text{Cher}}}{P_{\mathrm{vac}}(k)} = c_s \geff^2\dfrac{\pi\,n_A}{6 M^2\,H}|c_sk\eta_n|^2 +{\cal O}(|c_sk\eta_n|^4).
\ee
Let us make a few general remarks about these results. First, note that the emitted power vanishes in the $k\eta_n\to 0$ limit. This is in agreement with causality, which prevents the excitation of the super-Hubble modes at the instant of the particle production. As was observed above, for shorter modes that lie below the UV cut-off, i.e. $(-k\eta_n)<\sqrt{M/H}$, the power spectrum of the emitted inflaton asymptots to a constant plus tiny modulations. 

%


Imagine that the periodic particle productions occur with the frequency $\omega$.  
Due to the Cherenkov emission, each event excites the following window of momenta
\be
\label{range}
\dfrac{1}{|\eta_n|}\lesssim k<\dfrac{1}{|\eta_n|}(M/H)^{1/2}\,.
\ee
Since $\frac{\eta_{n+1}}{\eta_{n}}=e^{-2\pi\frac{H}{\omega}}$, any two subsequent events have overlapping range of excited modes iff
\be
\label{freqlim}
\dfrac{\omega}{H}>\dfrac{4\pi}{\ln M/H}\,,
\ee
and otherwise each excited mode exclusively is attributed to an event. In the high frequency limit, namely when $\omega\gg H$, 
the sum over $\eta_n$ can be replaced with an integral, i.e.
\be
\sum_{n} \rightarrow \dfrac{\omega}{2\pi\, H}\int \dfrac{d\eta_n}{\eta_n}\,,
\ee
The integral (sum) must be cut for $|\eta_n| > \frac{1}{k} (M/H)^{1/2}$, because of \eqref{range}. Altogether, this results in 
\bea
\label{Power:summed}
&&\frac{\expect{\pi_c(\k,0) \pi_c(-\k,0)}'_{\text{ppo}}}{P_{\text{vac}}(k)} \simeq 8 c_s \geff^2 \dfrac{H^2}{M^2} N_{A} \,\left(1-\gamma_{\mathrm{E}}+\mathrm{Ci}(\mu)-\log \mu -\dfrac{\sin \mu}{\mu} \right),\qquad \mu \equiv 2c_s (M/H)^{1/2}
\eea
in which $N_{A}$ is the total number of created $A$ particles per Hubble patch per Hubble time, i.e.  
\be
N_{A}=\dfrac{n_{A}}{H^3}\dfrac{\omega}{H}.
\ee
We are interested in the regime where the back-reaction of the created massive particles on the background evolution is negligible. Thus we ask the energy density of $A$ particles, which in the high frequency limit saturates to 
\be
\text{high frequency:}\qquad \rho_{A} \sim \frac{1}{3}\dfrac{\omega}{H}\,Mn_{A} \,,
\ee
to be much smaller than the $3M_p^2H^2$
\footnote{Technically, $\dot{\rho}_A$ should be small with respect to $\dot{\phi}_b\sim 3M_p^2|\dot{H}|$ too. But, in the high frequency limit, this does not provide a more stringent bound.}
, i.e. 
\be
\text{high frequency:}\qquad\dfrac{\rho_A}{\rho_{\text{b}}}\sim \dfrac{M\,n_A \omega}{9 \mpl^2 H^3}\ll1\,,
\ee
and that puts an upper bound on $N_A$, i.e. 
\be
N_A\ll 9\Big(\dfrac{M_p^2}{H\,M}\Big)\,.
\ee
The relative change in the power spectrum is
\be
\label{sigton2}
\dfrac{\Delta P}{P}=\dfrac{\langle \zeta^2\rangle_{\mathrm{Cher.}}}{\langle \zeta^2\rangle_{\mathrm{vac.}}}\sim {\cal O}(1)c_s\,g_{\text{eff}}^2\,(H/M)^2\,N_A\,,
\ee 
and due to the constraints on $N_A$ and assuming $g_{\text{eff}}\lesssim 1$, it is bounded from above, i.e.
\bea
\label{hfreq}
\label{Power:summed}
\text{high frequency:}\quad \dfrac{\Delta P}{P}\ll \dfrac{10^8}{\epsilon}\, (H/M)^3\,.
\eea
The upper bound can be huge and $\Delta P/P$ can be of order unity. However,
notice that in this case, namely the high frequency of particle production $\omega\gg H$, the Cherenkov effect only adds a scale-independent offset to the power-spectrum, which is indistinguishable from the vacuum fluctuations. Consequently, in this regime, the distinct features of the model should be sought  in the higher points statistics, or alternatively as we argue in the next section, in the patterns created by individual particles. 

In the low frequency limit, i.e. opposite to \eqref{freqlim}, the scale-dependence in the emission caused by each event, as depicted in Figure \ref{Cher-Spec:fig}, is distinguishable. In this limit, having negligible impact on the background demands $\dot{\rho}_A\ll M_p^2H\dot{H}$, which is stronger than $\rho_A\ll \rho_b$ used above. Hence
\be
\text{low frequency:} \qquad \dfrac{\dot{\rho}_A}{\dot{\rho}_b}\sim \dfrac{n_A\,M}{M_p^2\,H^2\,\epsilon}\ll 1\,,
\ee
which results in 
\be
\label{lfreq}
\text{low frequency:} \qquad \dfrac{\Delta P}{P}\ll 10^7\,(H/M)^3\,.
\ee
Therefore, by assuming $M\sim 100 H$, we see no obvious obstacle to the detectablity of the Cherenkov contribution to the power spectrum in scenarios with slower rates of particle production. 
\subsection{Real Space Radiation Profile of Rare Events}
\label{Cherenkov:Real-space}
Equation \eqref{pp:2pf} straightforwardly extends to higher order correlators (see Appendix A), such that a connected N-point function becomes
\bea
\nonumber
\dfrac{1}{\prod_{i=1}^N\, P^{\frac{1}{2}}_{\text{vac}}(k_i)}\Big\langle\Psi|\pi_c(\bfk_1,0)...\pi_c(\bfk_N,0)|\Psi\Big\rangle'_{\text{Cher}}&=&n_A\,a_n^3\,\int \dfrac{d^2\hat{q}}{4\pi}\times\\ 
\label{Npfeff}
&&\,\prod_{i=1}^N\,\Big[\geff\int_{\eta_n}^0 \dfrac{d\eta}{\eta\,H}\,\text{Im}(f_\pi '(k_i,\eta))\dfrac{1}{q_n}\,\exp(-i\bfk_i.\hat{q}\eta)\Big]\,,
\eea
and similar expressions can be derived for disconnected contributions as well. In effect, the above equation can be derived by writing $\pi_c$ as 
\be
\label{decomp}
\pi_c=\pi_c^{\text{QM}}+\pi_c^{\text{Cher}}\,,
\ee
such that (a)$\pi_c^{\text{Cher}}$ is a classical random field that corresponds to the Cherenkov radiation of a stochastic distribution of point-particles. Concisely, 
\be
\pi_c^{\text{Cher}}(\bfx)=\sum_{\bfx_n,\hat{q}}\, \pi_c^{\text{cl}}(\bfx;\bfx_n,\hat{q})\, Y(\bfx_n, \hat{q})\,, 
\ee
where $\pi_c^{\text{cl}}(\bfx;\bfx_n,\hat{q})$ is the radiation profile of a single particle with the initial position $\bfx_n$ and momentum $q\hat{q}$, and $Y(\bfx_n, \hat{q})$ is the same random variable as defined around \eqref{Ydis}.\,
 (b) $\pi_c^{\text{QM}}$ is the Heisenberg quantum operator corresponding to $\pi_c$ ,\,
(c) the initial state of the correlator should be replaced by the vacuum, and finally\, 
 (d) the two operators are un-correlated, i.e. $\la \pi_c^{\text{QM}}\pi_c^{\text{pp}}\ra=0$. The decomposition in \eqref{decomp} by itself suggests that the signal can be directly searched for inside the $\pi_c$ field (which is proportional to the curvature perturbation $\zeta$) rather than its correlation functions. This becomes especially useful for rare events: when the number of the particles per Hubble volume is not too large, therefore, the radiation pattern induced by  particles is distinguishable. 
 \begin{figure}[t!]
    \centering
    \begin{subfigure}[b]{0.45\textwidth}
        \centering
        \includegraphics[height= 2.5 in]{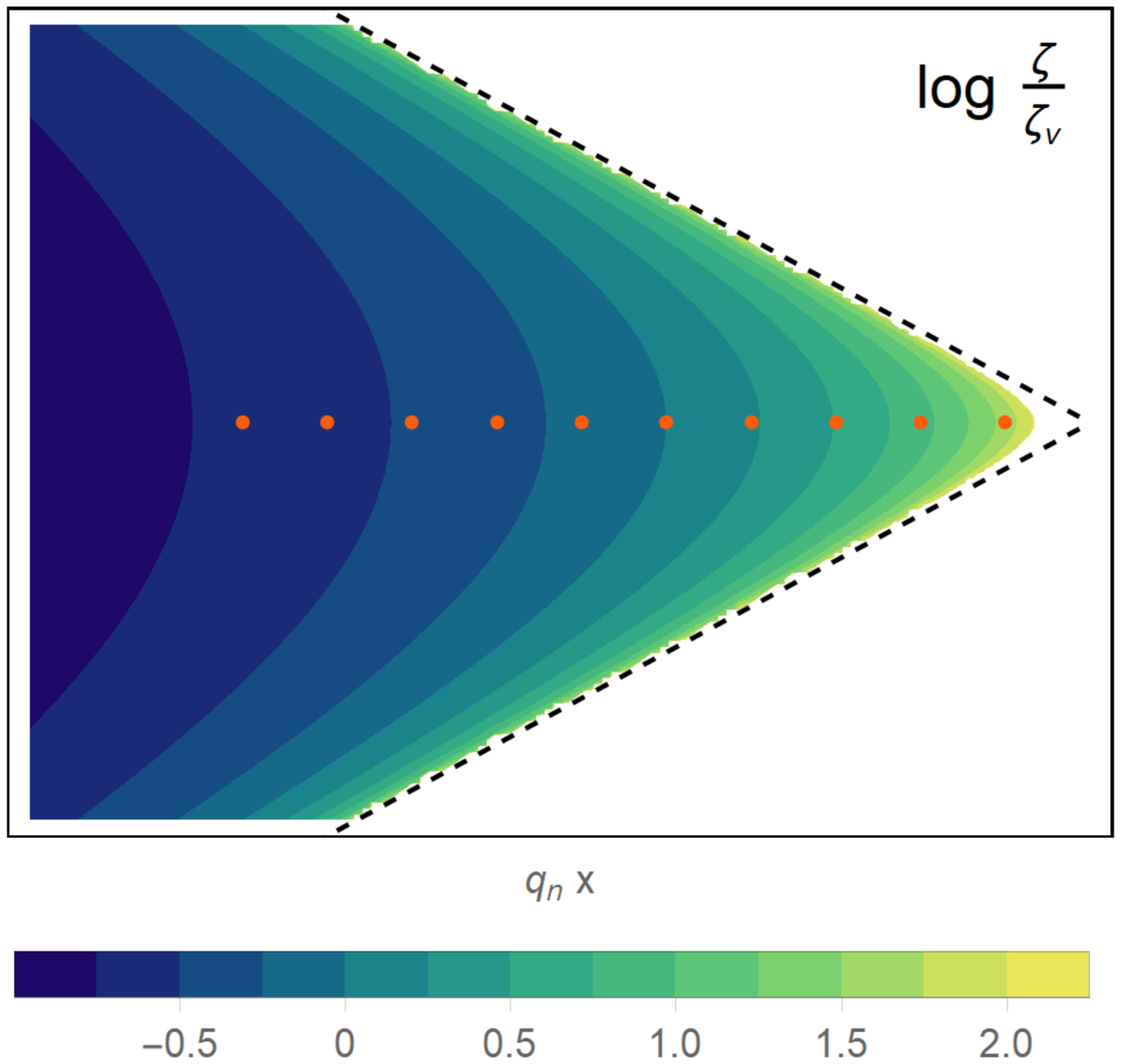}
        \end{subfigure}
    ~ 
    \begin{subfigure}[b]{0.45\textwidth}
        \centering
        \includegraphics[height= 2.5 in]{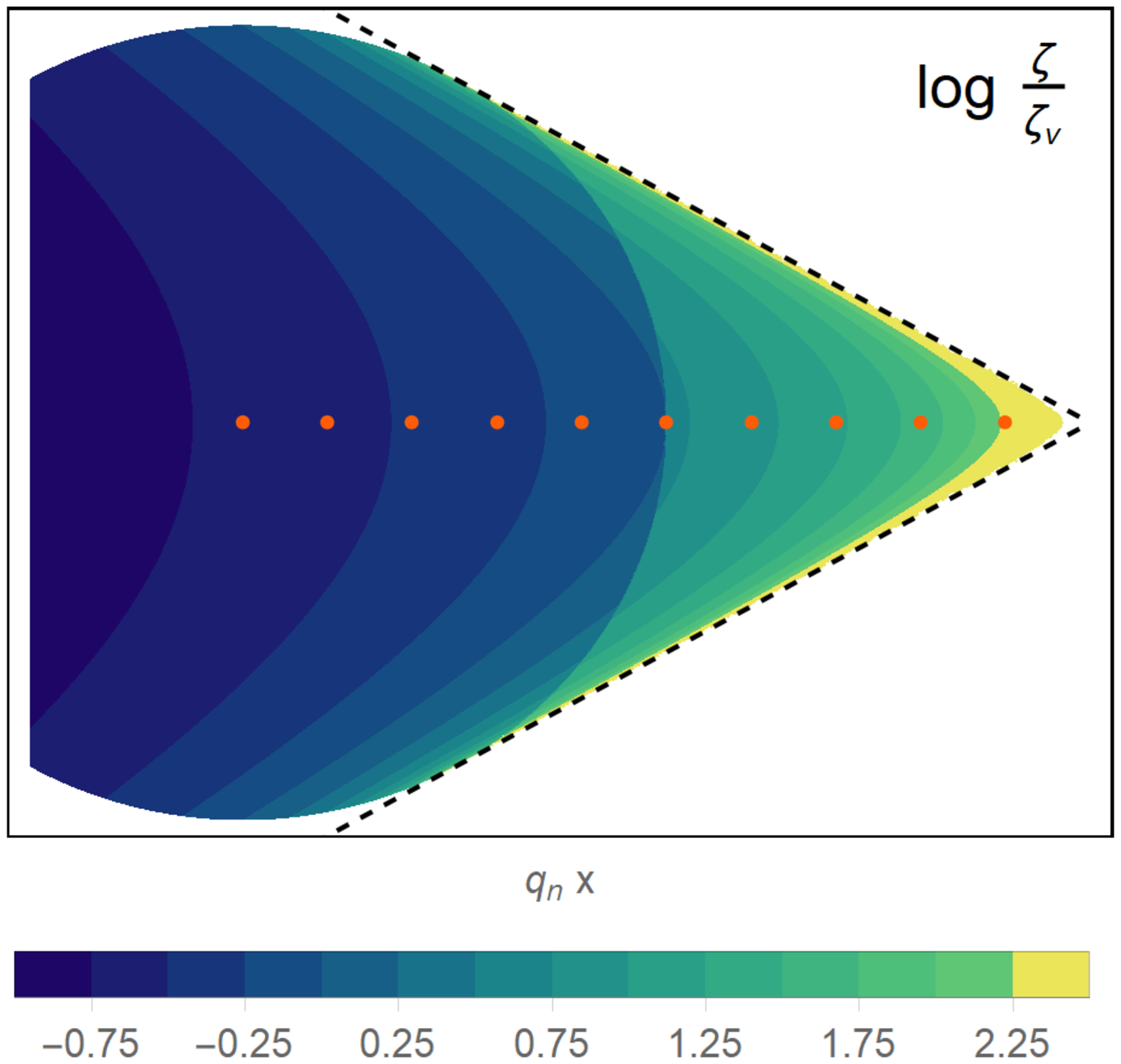}
         \end{subfigure}
    \caption{\label{machcone} A schematic illustration of the Cherenkov radiation imprint on the curvature perturbations, using Eq. \eqref{zeta-x:all}. Grayscale shows logarithm of the amplitude of the perturbations with some suitable normalization. In this figure we assume $c_s=0.7$. Red  point show the trajectory of the source $A$ while ``shock-front" is depicted in dashed black. (a) Far from production point $x=\eta_n$, (b) Including the production site $x=\eta_n$}
\end{figure}

Consider a single particle that embarks on its journey at $\eta_n$ from the initial position $\bfx_n$. By using \eqref{zeta-k:by-parts}, and knowing that the curvature perturbation on comoving slices $\zeta$ is related to $\pi$ via $\zeta=-H\pi$, we find the radiation in $\zeta$ induced by this single particle as
\bea
\label{zeteq}
\zeta(\x,0)&=& \dfrac{g_{\text{eff}}}{H q_n}\int_{\eta_n}^0 \dfrac{\md \eta'}{\eta'}  \int \frac{k^2 \md k}{(2\pi)^3}\, \mathrm{Im} f'_{\pi}(k,\eta')\,\zeta_{\mathrm{vac}}(k)\, \int \md \Omega_k\,e^{i\k\cdot (\x- \x_A)}\\ \nonumber
&=&g_{\text{eff}}~ \dfrac{A_s^{1/2}\,c_s^{1/2} }{q_n\,}\int_{\eta_n}^0 d\eta' \, \int_{0}^{\infty}dk\,\dfrac{1}{(2\pi)\Delta r}\sin(k\Delta r)\sin(c_s\,k\eta')\,,
\eea
where 
\be
\nonumber
\Delta \textbf{r}(\eta)=\x-\x_A(\eta)\,.
\ee
The expression inside the last integral above is the reminiscent of the retarded Green function in flat space and can be simplified as,
\be
\begin{split}
\int_{0}^{\infty}dk\,\dfrac{1}{(2\pi)\Delta r}\sin(k\Delta r)\sin(c_s\,k\eta')\,=\dfrac{1 }{(2\pi)\,\Delta r} \delta(c_s \eta'+ \Delta r)
 \,.
\end{split}
\ee
Without loss of generality, suppose that the worldline of the particle is $\x_A= \hat{q}\eta$. Using Eq.\eqref{zeteq}, then we find
\be
\label{zeta-x:all}
\zeta(\x,0)=
~\geff~\dfrac{A_s^{1/2}~c_s^{1/2}}{(2\pi)q_n}\sum_{\pm} \dfrac{1}{|c^2_s\eta_{\pm}+\hat{q}.\Delta \textbf{r}(\eta_\pm)|}\theta(\eta_\pm-\eta_n)\,, \qquad \text{iff}\qquad \hat{\bfx}.\hat{q}<-(1-c_s^2)^{1/2}\,,
\ee
and it vanishes otherwise, namely outside the Mach cone. Also
\be  
\eta_ \pm=\dfrac{1}{1-c_s^2}\Big(\bfx.\hat{q}\pm \Big((\bfx.\hat{q})^2-(1-c_s^2)r^2\Big)^{1/2}\Big)\,,
\ee
are the retarded times, which are simply two roots of the equation $c_s\eta+\Delta r(\eta)=0$. By simplifying \eqref{zeta-x:all} a step further, in the region where both $\eta_\pm$ are bigger than $\eta_n$, one finds
\be
\boxed{
\zeta(\x,0)=
~\geff~\dfrac{2A_s^{1/2}~c_s^{1/2}}{(2\pi)q_n}\,\dfrac{1}{\sqrt{(\bfx.\hat{q})^2-(1-c_s^2)r^2}}\,.
}
\ee
As expected, the curvature perturbation blows up towards the Mach cone, which  forms an approximate conical shock-wave with the apex angle, $2\phi = 2\sin^{-1}(c_s)$ (See Fig.\ref{machcone}).
Interestingly, this result coincides with its flat space counterpart in \eqref{cerenkov}, provided a change in the vertex from $\dot{\pi}A^2$ to $\pi\,A^2$, and inserting an appropriate coupling constant. This similarity between the dS and the flat space results is due to the following relation between the retarded Green function of the two
\be
\nonumber
\partial_\eta\, G_k(\eta,0)/\eta \propto G^{\text{flat}}_k(\eta,0)\,,
\ee
therefore, the redshift in the particle energy and the dilution in the wave function velocity in the dS space exactly cancel against each other such that the inflaton emission happens as if the background was non-expanding.
\footnote{Notice that this is only true if we look at $\zeta$ at the end of inflation, i.e. $\eta=0$.}

\begin{figure*}[t!]
\label{Cherenkov:fig}
    \centering
    \begin{subfigure}[b]{0.5\textwidth}
        \centering
        \includegraphics[height=4in]{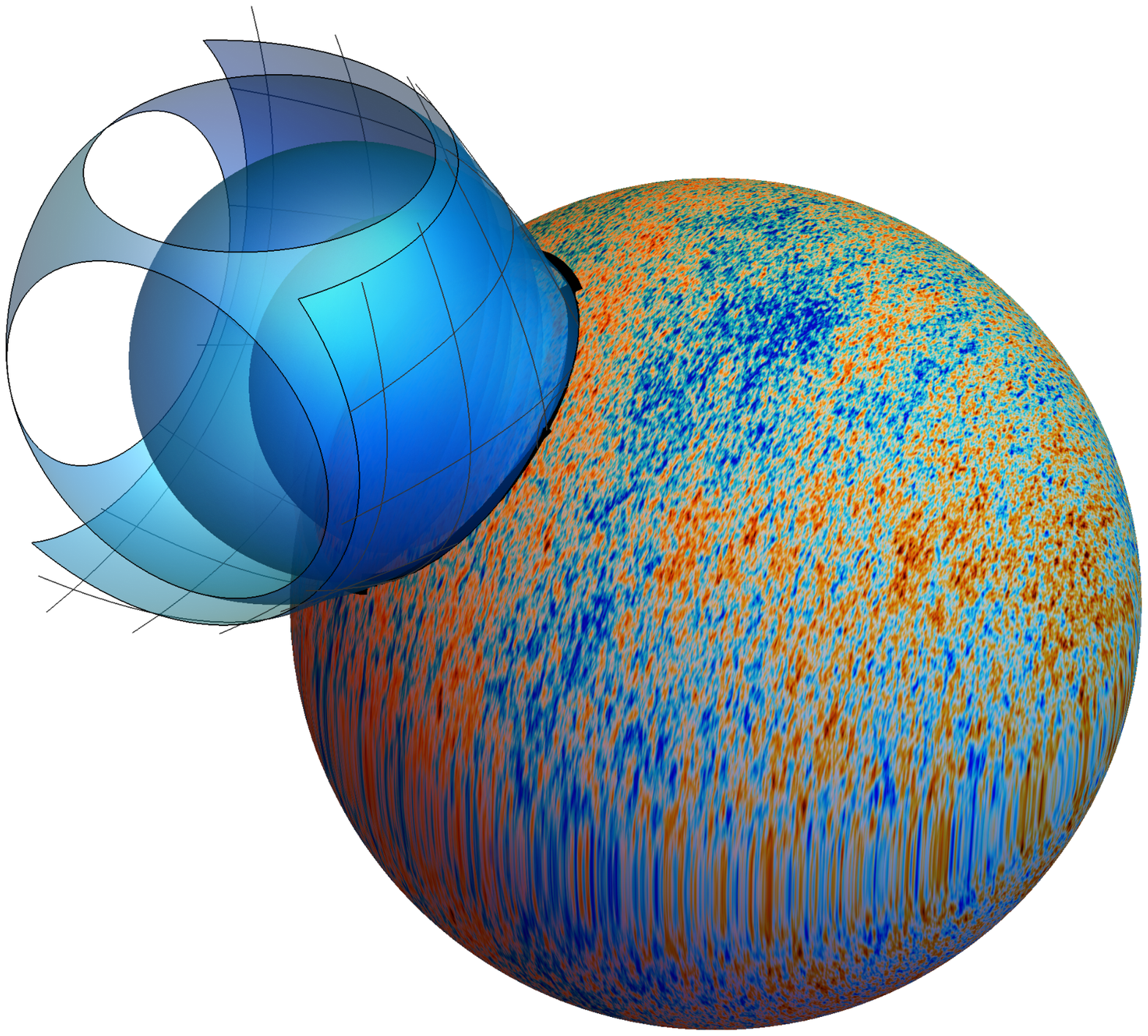}
        \caption{}
    \end{subfigure}%
    ~ 
    \begin{subfigure}[b]{0.5\textwidth}
        \centering
        \includegraphics[height=4in]{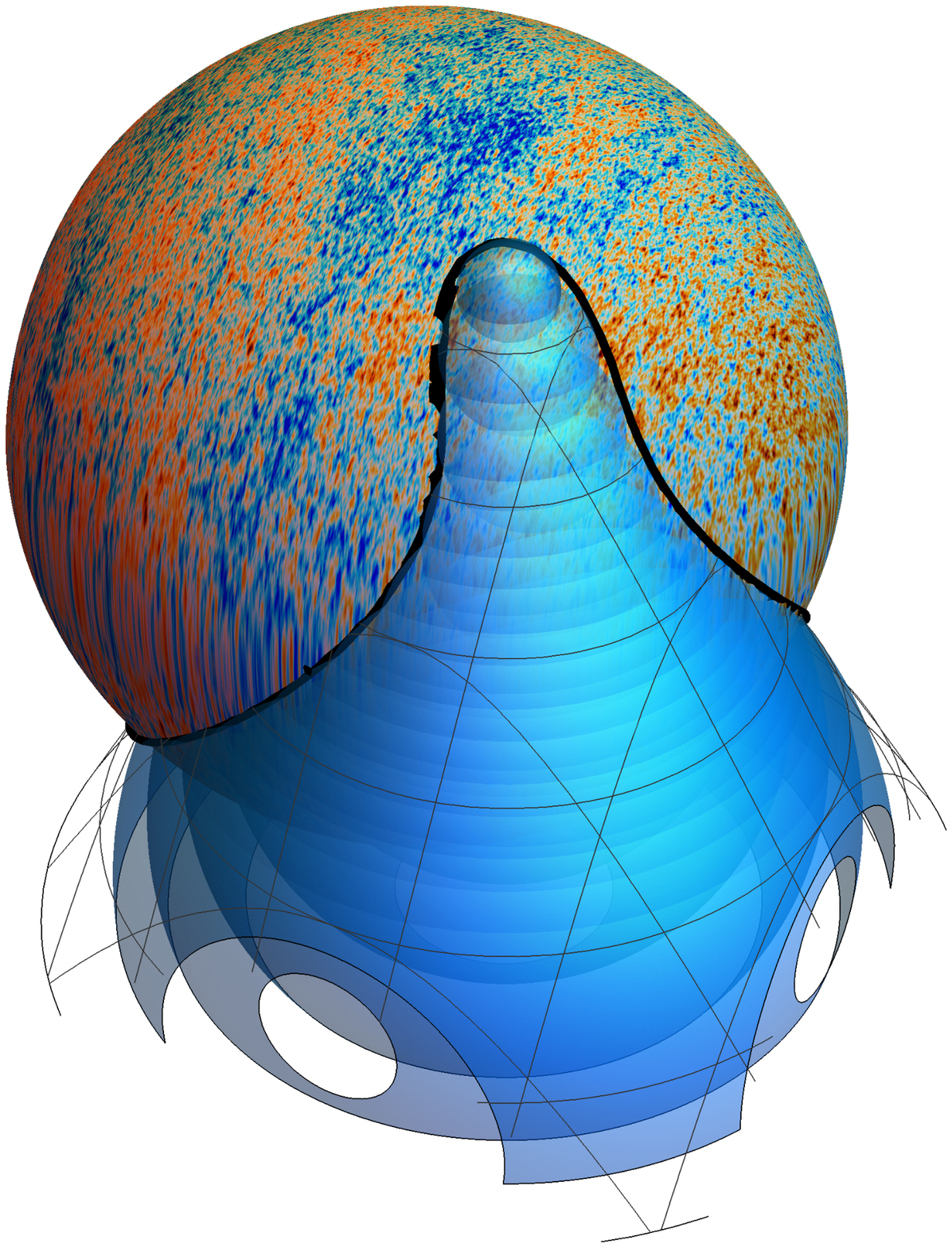}
        \caption{}
    \end{subfigure}
    \caption{\label{kinematic}A schematic illustration of the imprint of Cherenkov emission of inflatons on the CMB. Depending on how the shock-wave penetrate through the CMB sphere, there would be distinct feature. In general the intersection is a cone-sphere intersection. This can be divided into two major classes called on-axis and off-axis intersections depending on the heading of cone axis relative to the center of sphere. (a) The left shock-wave cone pierces the last scattering surface "on-axis" while (b) the right one is an "off-axis" example.}
\end{figure*}

The Cherenkov patterns induced on $\zeta$ would lead to distinctive  feautures on the temperature fluctuations of the Cosmic Microwave Background. In the present work, we are not to study those imprints in details, nevertheless, a rough picture can be easily obtained for large angle footprints: if the charachteristic scale of Cherenkov pattern, namely $L\sim \eta_n$, is bigger than the comoving size of the sound horizon at the recombination, the temperature fluctuations can be estimated as 
$\Theta(\hat{\mathrm{n}}) \propto \zeta(r_L\hat{\mathrm{n}})$ where $r_L$ denotes the radial coordinate to the last scattering surface. 
As long as the total number of particles per Hubble time per Hubble volume is not huge, the imprints of individual particles are resolvable. In the large angle approximation, the resulting pattern sharply peaks on the intersection of the particle Mach cone with the last scattering surface(see Fig. \ref{kinematic}).

\subsection{Estimating Non-Gaussianity}
\label{stat2}
In this section, we only give an estimate for non-Gaussianity originated from the Cherenkov emission. We differ a careful numerical evaluation of the bispectrum until a future work.  
Focusing on a single particle production event at $\eta_n$, a generic three-point function is given by
\bea
\dfrac{1}{\prod\limits_{i=1}^3\, P^{\frac{1}{2}}_{\text{vac}}(k_i)}\Big\langle\Psi\Big|\pi_c(\bfk_1,0)\pi_c(\bfk_2,0)\pi_c(\bfk_3,0)|\Psi\Big\rangle'_{\text{Cher}}=n_A\, a_n^3\,\geff^3\, (\dfrac{c_s}{2})^{3/2}\, 
\int \dfrac{d^2\hat{q}}{4\pi}\, \prod\limits_{i=1}^3\, \dfrac{1}{q_n}\,g(\bfk_i, \hat{q},\eta_n)\,.
\eea
This expression cannot be written in terms of elementary functions for a generic shape of the bispectrum, nevertheless, it is easy to see that it peaks around the flattened configuration, namely 
when all $\bfk_i$s are parallel and 
\be
\label{Mom-Const}
\sum_{i} p_i |\bfk_i| =0, \qquad p_i=\pm 1\,,
\ee
due to the momentum conservation. To see this, let us first perform the integration over $d^2\hat{q}$,
\bea
\label{d2q}
\int \dfrac{d^2\hat{q}}{4\pi} \prod_i \dfrac{ g(\k_i,\hat \q,\eta_n)}{q_n} &=&  \dfrac{1}{q_n^N}\Big( \prod_i \int_{\eta_n}^0 \mathrm{d}\eta_i ~ k_i^{1/2} \sin (c_s k_i \eta_i) \Big)\dfrac{\sin  \Big|\sum\limits_{i} \bfk_i \eta_i\,\Big|}{\Big|\sum\limits_{i} \bfk_i \eta_i\,\Big|}\\ \nonumber
&=&\left(\dfrac{\eta_n}{q_n}\right)^3\, \dfrac{1}{|k_t\eta_n|}\,\Big(\prod_{i=1}^3\, \int_{-1}^0\,k_i^{1/2}\, dx_i\, \sin (c_s|k_i\eta_n|x_i)\Big)\dfrac{\sin \Big(|k_t\eta_n|\Big |\sum\limits_i\,  (\dfrac{\bfk_i}{k_t})x_i\Big|\Big)}{\Big |\sum\limits_i\,  (\dfrac{\bfk_i}{k_t})x_i\Big|}\,,
\eea
where $k_t=\sum\,k_i$. The dimensionless integral is enhanced in the flattened limit, in which case the expression $\Big |\sum\limits_i\,  (\dfrac{\bfk_i}{k_t})x_i\Big|$ vanishes on a 2-d surface in $(x_1,x_2,x_3)$ space. From now on, we consider the flattened configuration only.
In the $|k_t\eta_n|\gg 1$ regime,
\footnote{We were unable to find an analytical estimate for  non-Gaussianity when $|k_t\eta_n|$ is of order unity. Nonetheless, after summing over all events  in the high frequency limit (i.e. $\omega\gg H$), one expects modes with $|k\eta_n|\gg 1$ to give the dominant contribution to the three-point function } 
one can approximate $\sin(|k_t\eta_n|\,x)/x$ with $\pi\delta(x)$. Consequently, we find
\be
\int d^2\hat \q \prod_i \dfrac{g(\k_i,\hat \q,\eta_n)}{q_n} = \pi\,\left(\dfrac{\eta_n}{q_n}\right)^3\, \dfrac{1}{|k_t\eta_n|}\,\Big(\prod_{i=1}^3\, \int_{-1}^0\,k_i^{1/2}\, dx_i\, \sin (c_s|k_i\eta_n|x_i)\Big)\delta \Big(\,\sum\limits_i\,p_i (\dfrac{k_i}{k_t})x_i\,\Big)\,.
\ee
Integrating over $dx_i$ gives us
\bea
\int d^2\hat \q \prod_{i=1}^{3} \dfrac{g(\k_i,\hat \q,\eta_n)}{q_n}=\dfrac{\pi\,|\eta_n|}{c_s\,q_n^3}\dfrac{1}{8\,\sqrt{k_1k_2k_3}}\Big(k_2\cos(2c_s|k_3\eta_n|)+k_3\cos(2c_s|k_2\eta_n|)-k_1\Big)+{\cal O}(\dfrac{1}{|k_{2,3}\eta_n|^3})\,.
\eea
where without loss of generality, we have assumed $k_1=k_2+k_3$. Summing over all events we find
\bea
\dfrac{\langle\zeta(\bfk_1)\zeta(\bfk_2)\zeta(\bfk_3)\rangle'_{\text{Cher}}}{\prod\limits_{i=1}^{3}\,P^{1/2}_\zeta(k_i)|_{\text{vac}}}&=& \pi \sqrt{\dfrac{c_s}{8}}\, \geff^3\, \dfrac{1}{M^3}\, \dfrac{1}{\sqrt{k_1k_2k_3}}\times \\ \nonumber
&&\, \sum\limits_{n}|\eta_n|\,\Big(k_2\cos(2c_s|k_3\eta_n|)+k_3\cos(2c_s|k_2\eta_n|)-k_1\Big)\,.
\eea
In the higher frequency limit, $\sum_n$ can be replaced with $\dfrac{\omega}{2\pi}\displaystyle \int\,\dfrac{d\eta_n}{H\eta_n}$, while as was discussed before, the time integral should be cut at $k_1^{-1}\,\sqrt{M/H}$. Consequently
\bea
\label{cherbi}
&&B_{\zeta}(k_1,k_2,k_3)=\sqrt{\dfrac{c_s}{128}}\,\geff^3\,\left(\dfrac{H}{M}\right)^3\,N_A\,(2\pi^2\,A_s)^{3/2}\,\dfrac{1}{k_1^2\,k_2^2\,k_3^2}\,\times \\ \nonumber
&&\Big(\dfrac{k_2}{2c_s\,k_3}\sin(2c_s\,(\frac{k_3}{k_1})\sqrt{M/H})+\dfrac{k_2}{2c_s\,k_3}\sin(2c_s\,(\frac{k_3}{k_1})\sqrt{M/H})-\sqrt{M/H}\Big)\,.
\eea
The exact oscillatory behavior displayed above cannot be seriously relied upon inasmuch as they are computed in the $M/H\gg 1$ limit and by using a sharp cut on $\eta_n$. Yet, the possibility of having observable oscillations in the flattened bispectrum, especially in the $c_s\ll 1$ regime, is interesting in light of the Planck recent data, which seems to mildly prefer an oscillatory flattened non-Gaussianity \cite{Ade:2015ava}. Capturing the right oscillatory behaviors in the bispectrum 
calls for a rigorous numerical evaluation in our set up, which we leave to a future work. Regardless, we can give an estimate for $f_{NL}$ by focusing on the third term appearing in \eqref{cherbi}. In the limit that the contribution from the Cherenkov effect to the two point function is suppressed with respect to the pure vacuum fluctuations, i.e. when $\Delta P/P$ in \eqref{sigton2} is small, we have
\be
f_{NL}\sim  \dfrac{B_{\zeta}(k,k,k)}{4 P^2_{\zeta}|_{\text{vac}}(k)}\sim \dfrac{\sqrt{c_s}}{64\pi}A_s^{-1/2}\,\geff^3\,(H/M)^{\frac{5}{2}}\,N_A\sim \dfrac{1}{\sqrt{c_s}}\,\dfrac{A_s^{-1/2}}{64\pi}\, \geff\, (H/M)^{1/2}\, \dfrac{\Delta P}{P}\,.
\ee
Therefore, sizeable amount of non-Gaussianity in the flattened shape is achievable, especially for cases with highly reduced speed of sound.  

Conversely, if the power spectrum is dominated by the Cherenkov radiation, $f_{NL}$($\sim B_\zeta/4P^2_{\text{Cher}}(k)$) will be suppressed by $1/N_A$. This is a simple manifestation of the centeral theorem---the sum of the radiation profiles of independent stochastic sources becomes extremely gaussian as $N_A$ goes to infinity.
\section{Conclusion}
\label{Conclusion:Sec}
In this work, we studied the footprints of supersonic particles during inflation, considering them as the decay products of yet another previously generated massive sector. We showed that as long as the initial phase space occupation number is small, one can effectively replace the initial non Bunch-Davies vacuum of the relativistic sector with a classical random distribution of point-particles and compute the radiation profile of each particle by making use of the point-particle effective action formalism. 

Regarding the statistical properties of the emitted scalar fluctuations, we computed the power spectrum and the (flattened) bispectrum of curvature perturbations due to the Cherenkov effect. Within the parameter space in which our analytical results pertains, detectable alteration to the typical inflationary power spectrum can be easily obtained, characterized by  $c_s$-dependent oscillatory features. Conversely, even in the case of negligible contribution to the power spectrum, we showed that sizeable amount of non-Gaussianity in the flattened shape can be obtained. We could not exclude the possibility of having appreciable resonant non-Gaussianity in the flattened shape before performing a numerical analysis, which we left to address elsewhere. 

When the number of created particles per Hubble patch is small, the radiation profile of individual particles are distinguishable and can be computed. The result, which is exactly the same as the conic structure in the flat space Cherenkov effect, would be the most distinctive feature of our model. 

Our study can be extended in a few directions: (a) First and foremost, it is interesting to confront the predictions of our set up with the CMB data. Of our particular interest is searching for rare events, for which the intersection of the shockwave and CMB surface leaves distinctive patterns on the temperature anisotropies. (b) As a complementary investigation, it would be interesting to consider spinning supersonic particles, as well as other interactions between the point-particle and the Nambu-Goldstone mode of the broken time translation within an EFT framework. (c) It is necessary to perform a numerical computation of non-Gaussianity for general shapes and in particular for the flattened configuration, in order to accurately account for potentially observable resonances in the bispectrum. (d) In this paper, we neglected the interference effect, as they are suppressed with at least a factor of $H/M$. Nonetheless, if we allow $H/M$ to exceed the occupation number $\beta$, then the leading order contribution to N-point functions comes from ${\cal O}(\beta)$ interference effects. In such regimes, it would be compelling to find similar $c_s$-sensitive features in the quantum contributions to the inflaton correlation functions. 
\section{Acknowledgement}
We are greatly indebted to Mehrdad Mirbabay for his initial collaboration and stimulating discussions through the development of this project. We would like to thank Matteo Fassiello, Garrett Goon, Blake Sherwin and David Wands for fruitful discussions. We are grateful to Hassan Firouzjahi and Enrico Pajer for commenting on the manuscript. AAA acknowledges support of school of physics of IPM,where he is a part-time member. AAA is thankful to the International Centre for Theoretical Physics (ICTP)  for hospitality whilst parts of this project were in progress.   
\appendix
\section{Point-Particle Description Vs. In-In Formalism}
\label{ppinin}
\subsection{Flat Space}
This appendix aims at justifying the point-particle approximation set forward in \eqref{Npt} and \eqref{classical}, for an arbitrary N-point function. We sketch the proof for a wave-packet like initial state as in \eqref{packet}, however, the logic similarly justifies the point particle approximation for more realistic initial states such as squeezed states of the form \eqref{sqst}, albeit with $\beta^2\ll 1$. 

Consider the in-in expression for an N-point function of $\vpi$,
\be
\label{ININVSPP}
\Big\la \Psi\Big|\,\bar{T}\Big(\exp(i\int_{\eta_n}^0\,d\eta\,H_I(\eta))\Big)\,\vpi_{\bfk_1}(0)\,...\,\vpi_{\bfk_N}(0)\,T\Big(\exp(-i\int_{\eta_n}^0\,d\eta\,H_I(\eta))\Big)\,\Big|\Psi\Big\ra.
\ee
The leading order connected contribution
 to the expression is of order $g^N$, and at the same time disconnected terms of order $g^M$ with $M<N$ are present. Hereafter we disregard the subleading terms of order $g^M$ with $M>N$. A sample ${\cal O}(g^M)$ piece that incorporates $R>0$ and $M-R>0$ number of interactions from the anti-time order and time-order operators, respectively, takes the following form 
\bea
\label{IMR}
&&{\cal I}_{M,R}=(-1)^{M-R}\,(\dfrac{ig}{2})^M\,\int_{\eta_n}^0\,d\eta_1 \,\int_{\eta_n}^{\eta_1}\,d\eta_2...\int_{\eta_n}^{\eta_{R-1}}d\eta_R\,\int_{\eta_M}^{0}d\eta_{R+1}\,...\,\int_{\eta_n}^{\eta_{M-1}}d\eta_M\,\int \prod_{i=1}^M\,d^3x_i\\ \nonumber
&&\Big\la \Psi|\,\underbrace{\vpi(x_1)A(x_1)^2\,...\,\vpi(x_R)A(x_R)^2}\,\vpi_{\bfk_1}(0)...\vpi_{\bfk_N}(0)\, \underbrace{\vpi(x_{R+1})A(x_{R+1})^2\,...\vpi(x_N)A(x_N)^2}\,|\Psi\Big\ra\,,
\eea
For $M<N$, \eqref{IMR} is disconnected among the external $\vpi$ legs, i.e. we can write
\bea
\nonumber
&&I_{M,R}=(-1)^{M-R}\,(\dfrac{i\,g}{2})^M\,\int_{\eta_n}^0\,d\eta_1 \,\int_{\eta_n}^{\eta_1}\,d\eta_2...\int_{\eta_n}^{\eta_{R-1}}d\eta_R\,\int_{\eta_M}^{0}d\eta_{R+1}\,...\,\int_{\eta_n}^{\eta_{M-1}}d\eta_M\,\int \prod_{i=1}^M\,d^3x_i\\ \nonumber
&&\qquad\quad\times\la 0|\vpi_{\bfk_{M+1}}(0)...\,\vpi_{\bfk_N}(0)|0\ra\,\la 0|\vpi(x_1)\,..\,\vpi(x_R)\vpi_{\bfk_1}(0)..\vpi_{\bfk_{M}}(0)\vpi(x_{R+1})..\vpi(x_M)|0\ra\,\,\la\Psi|A^2(x_1)\,...A^2(x_M)|\Psi\ra\,\\ \nonumber
&&\qquad +\Big(\frac{N!}{M!(N-M)!}-1\Big)\,\text{perm}\,\\ \nonumber
&&\qquad\quad=(-1)^{M-R}\,(i\,g)^M\,\int_{\eta_n}^0\,d\eta_1 \,\int_{\eta_n}^{\eta_1}\,d\eta_2...\int_{\eta_n}^{\eta_{R-1}}d\eta_R\,\int_{\eta_M}^{0}d\eta_{R+1}\,...\,\int_{\eta_n}^{\eta_{M-1}}d\eta_M\,\la 0|\vpi_{\bfk_{M+1}}(0)...\,\vpi_{\bfk_N}(0)|0\ra\,\\ \nonumber
&&\qquad\quad\times\,\Big[\Big(\prod_{i=1}^M\,\vpi_{\bfk_i}(0)\prod_{j=1}^R\vpi_{\bfk_j}(\eta_j)\prod_{l=R+1}^M\,\vpi^*_{\bfk_l}(\eta_l)\Big)\la \Psi|A^2(\bfk_1,\eta_1)...A^2(\bfk_M,\eta_M)|\Psi\ra+\,(M!-1)\,\,\text{perm}\Big]\\ 
&&\qquad\quad+\Big(\dfrac{N!}{M!(N-M)!}-1\Big)\,\text{perm}\,.
\eea
where $A^2(\bfk,\eta)$ is the Fourier transformation of $A^2(\bfx,\eta)$.
Now a great simplification occurs if we focus only on oscillatory terms in $\la \Psi|A^2(\bfk_1,\eta_1)...A^2(\bfk_M,\eta_M)|\Psi\ra$ that resonate in the soft limit $|\bfk_i|\to 0$ and enhance the in-in integral. This approximation amounts to the following Wick contractions inside the mentioned bracket 
\bea
&&\Big\la \Psi|A^2(\bfk_1,\eta_1)...A^2(\bfk_M,\eta_M)|\Psi\Big\ra \sim \int_{\bfq}\int_{\bfq'}\,{\cal A}^*(\bfq')\,{\cal A}(\bfq)\,\times\\ \nonumber
&&2^M\, \Big\la \contraction[2ex]{0|}
{a_{\bfq'}}{\,\int_{\bfk'_1}\,}{A(\bfk'_1,\eta_1)}
0|
a_{\bfq'}\,\int_{\bfk'_1}\,A(\bfk'_1,\eta_1)
\contraction[2ex]{}{A(\bfk_1-\bfk'_1,\eta_1)}{\,\int_{\bfk'_2}}{A(\bfk'_2,\eta_2)}
A(\bfk_1-\bfk'_1,\eta_1)\,\int_{\bfk'_2}A(\bfk'_2,\eta_2)
... 
\contraction[2ex]{}{A(\bfk'_{M-1},\eta_{M-1})}
{\int_{\bfk'_M}}{A(\bfk'_M,\eta_M)}
A(\bfk'_{M-1},\eta_{M-1})
\int_{\bfk'_M}A(\bfk'_M,\eta_M)
\contraction[2ex]{}{A(\bfk_M-\bfk'_M,\eta_M)}{}
{a^\dagger_{\bfq}}
A(\bfk_M-\bfk'_M,\eta_M)a^\dagger_{\bfq}
|0\Big\ra\,.
\eea
Assuming that $|\sum_i \bfk_i|<\Delta p$, we arrive at the following substitution 
\be
\label{A2A2fl}
\Big\la \Psi\Big|A^2(\bfk_1,\eta_1)...A^2(\bfk_M,\eta_M)\Big |\Psi\Big\ra\to \dfrac{1}{p^N}\exp(-i\bfk_1.\hat{p}\eta_1)...\exp(-i\bfk_M.\hat{p}\eta_M)\,.
\ee
For $|\sum_i\bfk_i|>\Delta p$, $I_{M,R}$ undergoes an exponential suppression originating from the tail of ${\cal A}(\bfq)$. Moreover, the above approximation only holds when the modes are soft, i.e. $k_i^2|\eta_n|\ll p$, such that $k_i^2$ order terms appearing inside the phases of the oscillations ( e.g. $\exp(i|\bfk_i+\p|\eta-ip\eta)$) can be neglected.  
Plugging the above simplified expression inside \eqref{IMR} and realizing that all time (anti-time) ordered integrals can now be transformed into ordinary integrals, we arrive at the following expression 
\bea
\sum_{R=0}^{M}\,I_{M,R}&=&\la 0|\vpi_{\bfk_{M+1}}(0)...\,\vpi_{\bfk_N}(0)|0\ra\times \prod_{i=1}^M\,-g\int_{\eta_n}^0\,d\eta \,\text{Im}\Big[\vpi_{k_i}(0)\vpi_{k_i}(\eta)\Big]\,\dfrac{1}{p}\exp(-i\bfk_i.\hat{p}\eta)\,\\ \nonumber
&&+(\frac{M!}{N!(N-M)!}-1)\qquad \text{perm}\,. 
\eea
And this completes our proof for validity of \eqref{Npt}. 
\subsection{dS space}
Here we elaborate on the validity of the point particle approximation in the scenario that we considered in \ref{Che-Rad:Inflation}. We sketch the proof only for the two point function, however, the argument extends to higher point correlation function in the same manner as did occur in our flat space toy example above.

Along the lines of  \cite{Flauger:2016idt}, suppose that in an event of particle production, taking place at $\eta=\eta_n$, an immense number of non-relativistic particles with mass $M$ becomes non-adiabatically generated. Right after the event, the quantum state of the universe can be taken as the direct product of the inflaton vacuum state and a squeezed state of $\chi$ particles, i.e.
\be
|\Phi\ra= \,{\cal N}\,\exp\Big(\int_\bp\dfrac{\beta_p}{2\alpha_p}\,a_\chi(\bp)^\dagger\,a_\chi(-\bp)^\dagger\Big)\,|0\ra_\chi \otimes |0\ra_{\pi}\,,
\ee
where $a_{\chi}$ is the annihilation operator assigned to the massive particles $\chi$, and ${\cal N}$ is an inconsequential normalization factor. We will be interested in the regime of the validity of the point-particle approximation, therefore $\beta_q$ is supposed to describe a distribution of localized non-relativistic particles, meaning $\beta$ should have a sharp peak around $q\sim 0$ with a width $\Delta q$ much smaller than the rest mass $M$. In addition, we must have $\beta_q\ll 1$.
\footnote{Obviously, in the opposite regime, i.e. $\beta_q\gg 1$, the state is yet extremely classical, a property that will be manifest in the basis of the field operator $\phi(\bfx)$ and not in the particle description.}
 As we mentioned earlier, $\chi$ particles are supposed to quickly decay into a pair of $A$ particles such that the Schrodinger state develops into the following 
\be
\Big|\Psi\Big\ra\sim {\cal N}\Big(\,1+\int_{\bp}\int_{\bfQ}\int_{\bfQ'}\dfrac{\beta_p}{2}{\cal A}(\bfQ,\bp-\bfQ)\,{\cal A}(\bfQ',-\bp-\bfQ')\,a^\dagger_A(\bfQ)a^\dagger_A(\bp-\bfQ)a^\dagger_A(\bfQ')a^\dagger_A(-\bp-\bfQ')\Big)\Big|0\Big\ra\,,
\ee
where ${\mathcal A}(\bp,-\bp+\bfq)$ is the resonance amplitude corresponding to the decay of $\chi$ particles (with the momentum $\bp$) into two a pair of $A$ particles (with momenta $\bfq$ and $\bp-\bfq$). This amplitude may be taken as a
Breit-Wigner distribution, i.e.
\be
|{\mathcal A}(\bfk,-\bfk+\bfq)|^2 \propto \dfrac{1}{(q^2-M^2/4)+\Gamma^2 M^2/4}\,,
\ee
with $\Gamma\gg H$ standing for the decay rate. Nevertheless, the details of the decay amplitude ${\cal A}$ will be immaterial for our computations--- any momentum dependence will be integrated over and simply gives the total number density of $A$ particles $n_A$. 
In addition, the initial state $|\Psi\ra$ should be normalized such that 
\be
\la \Psi|a_A^\dagger(\p)a_A(\p)|\Psi\ra=4\,a_n^3\,n_{\chi}=4\int_{\bfq}|\beta_q|^2\,, \quad a_n=-\dfrac{1}{\eta_n\,H}\,,
\ee
where $n_\chi$ is the physical initial number density of $\chi$ particles, which is one quarter of $n_A$.  Let us also  define the initial comoving momentum of generated $A$ particles as $q_n$. We must also have $\frac{M}{2}\sim |q_n\eta_n|\,H$, with an uncertainty of order $\text{max}(\Delta q, \sqrt{\Gamma M})$.  
\footnote{We neglect the deviation of $\chi$'s from on-shell which allow for A particles to have slightly different initial momenta. As a matter of fact, the width of the distribution ($\Gamma$ in this case) does not have any influence on suppressing the radiation in the UV, a role that is instead played by the UV cut-off, $|k_{\text{UV}}\eta_n|\sim M/m_A$}

The scenario that we described in section \ref{Che-Rad:Inflation} differs from the flat space toy example not only due to the cosmic expansion, but also because of the different vertex, i.e.  
\be
{\cal S}_{\text{int}}=g_{\mathrm{eff}}\,\int\,d\eta\,d^3\bfx\dfrac{1}{\eta^3\,H^3}\,\pi_c'\,A^2\,.
\ee
Now lets us explicitly write down the leading order correction to the two point function due to the above interaction
\bea
\nonumber
&&\dfrac{1}{P_{\text{vac}}(k)}\,\Big\la \Psi\Big|\,\pi_c(\bfk,0)\pi_c(-\bfk,0)\Big|\Psi\Big\ra'_{\text{Cher}}=g_{\mathrm{eff}}^2\\ \nonumber
&&\times\Big\lbrace\dfrac{1}{2}\,\int_{\eta_n}^0\,d\eta\int_{\eta_n}^0\,d\eta_1\,\dfrac{1}{(\eta\,H)^3(\eta_1\,H)^3}\,f_\pi'(k,\eta)f_\pi^{'\,*}(k,\eta_1)\la\Psi|A^2(\bfk,\eta_1)A^2(-\bfk,\eta)|\Psi\ra\,\\ 
\label{2pfds}
&&\qquad -\text{Re}\,\int_{\eta_n}^0\,d\eta\int_{\eta_n}^{\eta}\,d\eta_1\, \dfrac{1}{(\eta\,H)^3\,(\eta_1\,H)^3}f_\pi '(k,\eta)f_\pi '(k,\eta_1)\,\la\Psi|A^2(\bfk,\eta_1)A^2(-\bfk,\eta)|\Psi\ra\Big\rbrace +(\bfk\to -\bfk)
\eea 
Notice that if $A$ particles is lighter than $\sqrt{2}H$, this two point function is plagued by a secular IR growth in the $\eta,\eta'\to 0$ limit. There are a number of great advantages in assuming $m_A>\sqrt{2}H$. First, the aforementioned IR secular growth disappears. Second, the single field description for inflation would be valid, as the wave-function of $A$ decays rapidly outside the horizon. Third, the biggest contribution to the in-in integral comes from deep inside the past--- when the $A$ particle is relativistic, and where similar patterns of resonances that occurred in our flat space examples happens here as well. 

Three different types of contributions are present in $\la \Psi|A^2(\bfk,\eta_1)A^2(-\bfk,\eta)|\Psi\ra$ that deep in the past schematically look like
\be
\begin{split}
\nonumber
\la \Psi|A^2(\bfk,\eta_1)A^2(-\bfk,\eta)|\Psi\ra \sim \,& \int_{\bfQ_1}\int_{\bfQ}{\cal O}(\beta)~\eta\eta_1\exp(i\,Q_1\eta_1)\exp(i\,|\bfk-\bfQ_1|\eta_1)\exp(i\,\bfQ\eta)\exp(i\,|-\bfk-\bfQ|\eta)\\ \nonumber
+&\int_{\bfQ_1}\int_{\bfQ}{\cal O}(\beta^2)~\eta\eta_1\exp(i\,Q_1\eta_1)\exp(-i\,|\bfk-\bfQ_1|\eta_1)\exp(i\,\bfQ\eta)\exp(-i\,|-\bfk-\bfQ|\eta)\\
\label{A2A2psi}
+&\la 0|A^2(\bfk,\eta_1)A^2(-\bfk,\eta)|0\ra\, \la \Psi|\Psi\ra\,,
\end{split}
\ee
where the momenta of the relativistic particles inside the integral, namely $|\bfQ|, |\bfQ_1|$, are almost equal to $q_n=\frac{M}{2H\,|\eta_n|}$. Studying the last line which accounts for  loop contribution to the power spectrum is beyond the scope of current work (however, see \cite{Senatore:2009cf}), therefore we disregard it. 
By plugging \eqref{A2A2psi} into \eqref{2pfds} we observe that for a generic external momentum $k$, both ${\cal O}(\beta)$ and ${\cal O}(\beta^2)$ contributions are suppressed  by a factor of $(\frac{1}{q_n\eta_n})^2\sim H^2/M^2$. However, the ${\cal O}(\beta^2)$ terms can resonate for soft momenta defined through
\be
\label{reccut}
\exp(iQ\eta)\exp(-i|\bfk-\bfQ|\eta)\sim \exp(i\bfk.\hat{Q}\eta)\,
\ee
which is a  reasonable approximation for any momentum $k$ smaller than 
\be
k\lesssim k_{\text{UV}}=\sqrt{\dfrac{q_n}{\eta_n}}\sim \dfrac{1}{|\eta_n|}\left(\dfrac{M}{H}\right)^{1/2}\,.
\ee
Conversely, such resonance does not occur in ${\cal O}(\beta)$ contributions and they are always suppressed with a factor of $\frac{1}{q_n\eta_n}\sim H/M$ compared to resonating terms, therefore as long as $\beta$ is not too small (but still small enough to justify the point-particle description of the squeezed state), we can safely neglect ${\cal O}(\beta)$ terms. Making use of \eqref{reccut} then amounts to the following substitution 
\be
\label{dsaprx}
\la\Psi|\,A^2(\bfk,\eta)A^2(\bfk,\eta_1)|\Psi\ra\to\, 4\,(\eta_n\,H)^{-3}\,n_{\chi}\,H^4\int \dfrac{d^2\hat{q}}{4\pi}\,\eta^2\eta_1^2\,\dfrac{1}{q_n^2}\,\exp(-i\bfk.\hat{q}\eta)\exp(+i\bfk.\hat{q}\eta_1)\,,
\ee
which is the counterpart of \eqref{A2A2fl}. Inserting \eqref{dsaprx} back inside \eqref{2pfds} gives us the factorized form for the two point function that we had anticipated, i.e. 
\bea
\label{ininf}
&&\dfrac{1}{P_{\text{vac}}(k)}\,\Big\la \Psi\Big|\,\pi_c(\bfk,0)\pi_c(-\bfk,0)\Big|\Psi\Big\ra'_{\text{Cher}}=a_n^3\,\geff^2\,n_A\,\int\dfrac{d^2\hat{q}}{4\pi}\\ \nonumber
&&\left[\int_{\eta_n}^0 \dfrac{d\eta}{\eta\,H}\,\text{Im}(f_\pi '(k,\eta))\dfrac{1}{q_n}\,\exp(-i\bfk.\hat{q}\eta)\right]\,\left[\int_{\eta_n}^0 \dfrac{d\eta}{\eta\,H}\text{Im}(f_\pi '(k,\eta))\dfrac{1}{q_n}\,\exp(+i\bfk.\hat{q}\eta)\right]\,.
\eea

\end{document}